\documentclass[superscriptaddress,reprint,aps,pra,longbibliography]{revtex4-1}
\usepackage{graphicx}
\usepackage{amsmath}
\usepackage{amsfonts}
\usepackage{enumerate}
\usepackage{multirow}
\usepackage{makecell}
\usepackage[table,xcdraw]{xcolor}
\usepackage{times}
\usepackage[normalem]{ulem}
\usepackage[hidelinks]{hyperref}
\hypersetup{
    colorlinks,
    linkcolor={blue!80!black},
    citecolor={blue!80!black},
    urlcolor={blue!80!black}}

\begin{document}

\title{Quantum synchronization of few-body systems under collective dissipation}

\author{G. Karpat}
\email{goktug.karpat@ieu.edu.tr}
\affiliation{Faculty of Arts and Sciences, Department of Physics, \.{I}zmir University of Economics, \.{I}zmir, 35330, Turkey}

\author{\.{I}. Yal\c{c}\i nkaya}
\affiliation{Department of Physics, Faculty of Nuclear Sciences and Physical Engineering, Czech Technical University in Prague, B\v{r}ehov\'a 7, 115 19 Praha 1-Star\'e M\v{e}sto, Czech Republic}

\author{B. \c{C}akmak}
\affiliation{College of Engineering and Natural Sciences, Bah\c{c}e\c{s}ehir University, Be\c{s}ikta\c{s}, \.{I}stanbul 34353, Turkey}

\date{\today}

\begin{abstract}
We explore the environment-induced synchronization phenomenon in two-level systems in contact with a thermal dissipative environment. We first discuss the conditions under which synchronization emerges between a pair of two-level particles. That is, we analyze the impact of various model parameters on the emergence of (anti-)synchronization such as the environment temperature, the direct interaction between the particles, and the distance between them controlling the collectivity of the dissipation. We then enlarge the system to be composed of three two-level atoms to study the mutual synchronization between different particle pairs. Remarkably, we observe in this case a rich synchronization dynamics which stems from different possible spatial configurations of the atoms. Particularly, in sharp contrast with the two-atom case, we show that when the three atoms are in close proximity, appearance of anti-synchronization can be obstructed across all particle pairs due to frustration.
\end{abstract}

\maketitle

\section{Introduction}

Synchronization is a universal phenomenon that is fundamental to numerous topics both in natural and social sciences~\cite{Pikovsky2001, Osipov2007}. In general, synchronous behavior can be said to emerge in physical systems in two different manners: forced and spontaneous. When an external pacemaker drives a system and tends to impose its rhythm on it under suitable circumstances, there occurs forced synchronization also called entrainment. Nonetheless, synchronization might also manifest itself spontaneously, that is, solely as a consequence of the interaction between the subsystems, even without the existence of any external drive. In classical systems, the ubiquitous phenomenon of synchronization has been extensively explored in various different settings in last few decades~\cite{Arenas2008}. More recently, the study of synchronization has been extended to quantum regime and attracted considerable attention~\cite{Galve2017}.

In fact, synchronization could assume different definitions in the quantum domain depending on the context. On the one hand, certain studies consider the limit cycles of quantum mechanical systems to witness synchronous dynamical behavior, in close analogy with the classical theory of synchronization~\cite{Roulet2018,Roulet2018a,Jaseem2018,Koppenhfer2019}. On the other hand, some others focus on the transient quantum synchronization which actually describes the emergence of synchronized evolution between local observables of an open quantum system as it decays to its steady state~\cite{Giorgi2019}. Forced quantum synchronization has been examined in several different models including qubits coupled to dissipative resonators~\citep{Zhirov2008,Zhirov2009}, Van der Poll oscillators~\citep{Lee2013,Walter2014,Sonar2018} and spin-boson type models~\citep{Goychuk2006}. Besides, spontaneous quantum synchronization has been also well investigated, for instance, in systems of optomechanical arrays~\citep{Heinrich2011,Ludwig2013}, cold ions in microtraps~\citep{Hush2015}, Van der Poll oscillators~\cite{Lee2014,Walter2015}, harmonic oscillators~\citep{Giorgi2012,Manzano2013,Manzano2013a,Benedetti2016}, collision models~\cite{Karpat2019}, a pair of spins coupled to a common environment~\citep{Orth2010,Giorgi2013,Giorgi2016,Bellomo2017}, subject to incoherent pumping~\cite{Cabot2019-2}, and a dimer lattice with local dissipation~\cite{Cabot2019}.

Several techniques based on different merits have been proposed to identify the temporal correlations between the local dynamics of subsystems in order to witness the emergence of mutual quantum synchronization between them~\cite{Mari2013,Li2017}. In addition, certain well-known measures of quantum and total correlations (such as entanglement, quantum discord, and mutual information) contained in the global system have been analyzed as potential candidates to detect the occurrence of synchronous behavior between the local evolution of the subsystems~\cite{Galve2017}. Despite an apparent link between the onset of synchronization and the behavior of correlations in the global system in certain models, there exists no general connection between the two concepts. Indeed, exploiting the versatility of collision model framework, it has been recently shown that correlations in the global system play no relevant role for the dynamical synchronization of local observables~\cite{Karpat2019}. Lastly, relation of quantum coherence and synchronization has also been discussed in a bio-inspired vibronic dimer system~\cite{Jaszek2019}.

In this work, we study the dynamical establishment of mutual synchronization in a system of few two-level atomic particles due to their collective interaction with a dissipative thermal environment in the complete absence of an external drive. As a figure of merit, we use the Pearson correlation coefficient to quantify the degree of synchronization. We begin our analysis by discussing the necessary conditions for the spontaneous emergence of synchronization between expectation values of the local observables of a pair of two-level atomic systems. In particular, we recognize that the main ingredient responsible for the appearance of mutual synchronization is the collectivity of the interaction of the two atoms with the environment specified by inter-atomic distances. Even though the possibility of transient synchronization of a pair of atomic particles under dissipation has been discussed in relation with the collective phenomena of super- and sub-radiance for a zero temperature environment~\citep{Bellomo2017}, possible effects of a finite environment temperature on the emergence of synchronization under dissipation has not been yet examined. Here, we demonstrate that as the temperature of the environment increases, synchronized evolution between the local observables of the atoms cannot be maintained unless a certain degree of collective dissipation is ensured, which unveils the detrimental role of temperature for the onset of synchronization. Furthermore, despite a relatively large number of studies carried out on the relation between synchronization and correlations in the global system~\cite{Galve2017,Giorgi2019}, the degree of quantum coherence contained in the system has not been paid enough attention in this regard. Therefore, we explore the relation between the onset of spontaneous synchronization and the average residual quantum coherence contained in the two-atom system. Our findings show that there exists a connection between the two concepts at least in the case of a two-atom open system.

What is even more important is that we next turn our attention to the case of three two-level atoms. Almost all of the studies in the previous literature on the phenomenon of mutual quantum synchronization have typically limited their focus on a pair of two-level spins or oscillators undergoing open system dynamics, except for a very few number of papers which study networks of harmonic oscillators~\cite{Manzano2013, Manzano2013a}. As a consequence, in order to investigate the appearance of mutual synchronization between different bipartitions of a composite system, we consider three two-level atomic systems experiencing partial or full collective dissipation together in the absence of direct interaction among them. In this case, we find a quite rich synchronization dynamics showing that the nature of (anti-)synchronization between the local observables of different atom pairs is crucially dependent on the spatial configuration of atoms. For instance, we argue that when the spatial configuration of the three atoms is such that the degree of collectiveness of dissipation for all pairs of atoms is identical, there might occur a form of frustration in the synchronization dynamics and hence none of the atom pairs can exhibit fully anti-synchronized behavior in their observables. Relatedly, in striking contrast with the two-atom case, where (anti-)synchronization is robust independently of the choice of initial two-atom state, we show the strong dependence of the synchronization dynamics on the initial state of the three atom system. Hence, our results reveal that geometric effects such as frustration can play a decisive role on the nature of mutual synchronization even in the absence of direct interactions among the constituents of few-body quantum systems.

The remainder of the paper is organized as follows. In Sec.~\ref{model}, we introduce the open system model that we consider throughout the manuscript and the figure of merit characterizing synchronization phenomenon. Sec.~\ref{twoatoms} presents our results on the discussion of the conditions under which synchronization emerges for a pair of two-level systems, and the relationship between synchronization and   the residual coherence contained in the system. We display our results regarding the appearance of mutual synchronization for a system of three particles in Sec.~\ref{threeatoms}, and finally we conclude in Sec.~\ref{conclusion}.

\section{The model and figure of merit}\label{model}

The model that we will consider in the rest of the work describes the interaction of two-level atomic systems with a quantized, thermal electromagnetic field environment. The self-Hamiltonian for the central two level systems is given by $H_s=\sum_{i=1}^N\omega_i\sigma_i^z$, where $\omega_i$ is the transition frequency between the energy levels of the $i^{th}$ atom and $\sigma$'s denote the usual Pauli matrices. Throughout this work, we set $\hbar=1$ and fix the units of the other parameters accordingly.

Moreover, we will assume that the two-level atoms have polarized dipole moments, $\mathbf{d_{eg}}$, and that they can interact with each other through the dipole-dipole (exchange) interaction Hamiltonian $H_d=\sum_{i\neq j}^Nf_{ij}\sigma^+_i\sigma^-_j$ where $\sigma^{\pm}$ are the raising and lowering operators of the two-level atoms and $f_{ij}$ is the interaction strength. Taking into account the interaction with the thermal photons and looking at the reduced dynamics of two-level systems, one arrives at the following well-known, quantum optical master equation~\cite{BreuerPetruccione, Damanet2016}
\begin{equation}\label{me}
\dot{\rho} = -i\left[(H_s+H_d), \rho\right]+\mathcal{D}_-(\rho)+\mathcal{D}_+(\rho)=\mathcal{L}(\rho),
\end{equation}
with
\begin{align}
\mathcal{D}_-(\rho)=& \sum\limits_{i,j=1}^N\gamma_{ij}(\bar{n}+1)(\sigma_j^-\rho\sigma_i^+-\frac{1}{2}\{\sigma_i^+\sigma_j^-, \rho\}) \label{emission} \\ 
\mathcal{D}_+(\rho)= & \sum\limits_{i,j=1}^N\gamma_{ij}\bar{n}(\sigma_j^+\rho\sigma_i^--\frac{1}{2}\{\sigma_i^-\sigma_j^+, \rho\}). \label{absorbtion} 
\end{align}
The first term in Eq.~(\ref{me}) accounts for the unitary self-evolution and the dipole-dipole interaction between the two atoms. Following two terms introduce the interaction of the atoms with the environment. While the second term (see Eq.~(\ref{emission})) is responsible for the spontaneous and thermally induced emission, the third term (see Eq.~(\ref{absorbtion})) describes the thermally induced absorption process of the atomic system. The rates at which these processes take place are determined by the mean number of photons at the transition frequency of the atoms $\bar{n}=(\text{exp}(\beta\omega_i)-1)^{-1}$ with $\beta$ being the inverse temperature of the environment. Here, $\gamma_{ij}=\sqrt{\gamma_i\gamma_j}a(k_0r_{ij})$ where $\gamma_{i(j)}=\omega_{i(j)}^3 g$ with $g=d^2/3\pi\epsilon_0c^3$. Note that the positivity of the dynamics dictates that $a(k_0r_{ij})\leq 1$. Explicit forms of the parameters $f_{ij}$ and $a_{ij}$, which respectively control the strength of the exchange interaction between the two atoms and the degree of collectivity of the dynamics are given as~\cite{Bhattacharya2018, Damanet2016, Lehmberg, Stephen}
\begin{multline}
f_{ij}=\frac{3\gamma_0}{4}\left[(1-3\cos^2\alpha_{ij})\left(\frac{\sin\xi_{ij}}{\xi_{ij}^2}+\frac{\cos\xi_{ij}}{\xi_{ij}^3}\right)\right. \\ \nonumber 
\left. -(1-\cos^2\alpha_{ij})\frac{\cos\xi_{ij}}{\xi_{ij}}\right],
\end{multline}
\vspace{-0.4cm}
\begin{multline}
a_{ij}=\frac{3}{2}\left[(1-3\cos^2\alpha_{ij})\left( \frac{\cos\xi_{ij}}{\xi_{ij}^2}-\frac{\sin\xi_{ij}}{\xi_{ij}^3} \right)\right. \\ \nonumber 
\left. +(1-\cos^2\alpha_{ij})\frac{\sin\xi_{ij}}{\xi_{ij}} \right].
\end{multline}
Here, $\xi_{ij}=k_0r_{ij}$ is a dimensionless parameter characterizing the distance between the particles with $k_0=\omega_0/c$ and $r_{ij}=|\mathbf{r_{ij}}|=|\mathbf{r_i}-\mathbf{r_j}|$ is the relative positions of the $i^{\text{th}}$ and $j^{\text{th}}$ atom, and $\alpha_{ij}$ is the angle between $\mathbf{r_{ij}}$ and $\mathbf{d_{eg}}$.

The presented model has two extreme limits that depend on the spatial distance between the atoms considered inside the system. When the distance between the atoms is large as compared to the wavelength of the photons in the environment, $\xi_{ij}\gg 1$, the atoms couple to the environment individually. In the other extreme limit, $\xi_{ij}\ll 1$, the atoms are extremely close to each other, leading them to collectively interact with the environment. In terms of the collectivity parameter $a_{ij}$, the aforementioned cases can be characterized as $a_{ij}=0$ and $a_{ij}=1$, respectively. In the following section, we will discuss the effect of various model parameters on the synchronization dynamics of our main system.

Having detailed the open system model we plan to investigate in our work, let us now introduce the figure of merit for quantifying synchronization. In fact, one may directly examine the expectation values of the spin observables of system atoms to conclude whether they exhibit a synchronous behavior or not. In this way, the mutual synchronization is identified if the expectation values of individual atoms fluctuates in time in unison, i.e., if they oscillate with a fixed relative phase. The Pearson correlation coefficient $C_{xy}$ provides a convenient figure of merit for this characterization \cite{Galve2017}, which essentially measures the linear correlation between two given discrete variables $x$ and $y$ according to the formula
\begin{equation}
C_{xy}=
\frac{\sum_t(x_t-\bar{x})(y_t-\bar{y})}{\sqrt{\sum_t(x_t-\bar{x})^2}\sqrt{\sum_t(y_t-\bar{y})^2}}.
\label{eq:pc}
\end{equation} 
Here, the bars represent the variable averages over the data set indexed by $t$. The Pearson coefficient $C_{xy}$ varies within the interval $[-1,1]$ such that while $C_{xy}=0$ indicates that there is no linear correlation between two variables,  $C_{xy}~=~1 (-1)$ points out a positive (negative) linear correlation. In our work, the variables in question are the expectation values of the spin observables of the system atoms $s_1$ and $s_2$ in the $x$ direction, namely, $\langle\sigma_{s_1}^x\rangle$ and $\langle\sigma_{s_2}^x\rangle$, and we denote the corresponding Pearson coefficient by $C_{12}$. In the numerical simulations, we generate the expectation values as discrete samples covering the time interval under study. Then, we calculate the expression given in Eq. \eqref{eq:pc} over a sliding time window $\Delta t$ along the total simulation time to obtain a time dependent Pearson coefficient, and hence, to see how separate oscillations get phase-locked over time. Therefore, the summations and the averages that appear in Eq. \eqref{eq:pc} are taken over the time $\Delta t$, and the Pearson coefficient is calculated repeatedly for adjacent time windows until the whole simulation time range is covered. We also allow the adjacent time windows to overlap for an interval of $\delta t$ to prevent discrete jumps in the resulting $C_{12}$ curve (for further information, see the Appendix in \citep{Karpat2019}). In general, mutual synchronization is said to be established when $C_{12}$ does no longer change in time, i.e., when the expectation values oscillate in time with a constant relative phase difference. We will particularly study the cases where this phase difference is $0$ (synchronization) and $\pi$ (anti-synchronization). Other possible cases are referred in the literature as time-delayed synchronization \cite{Galve2017}. 

\section{Results}

Inside this section, we will present the outcomes of our investigation in three subsections. First, we begin by restricting our analysis to the case of two atoms that are undergoing collective dissipation. We discuss the consequences of the amount of detuning between the self-energies of the atoms and the strength of the direct interaction between them for the emergence of synchronization. Then, assuming resonant atoms, we explore the impact of the environment temperature and the spatial positions of the atoms on the appearance of synchronized behavior between them. Afterwards, we will examine the possible relation of average quantum coherence contained in the global system of two atoms to the degree of synchronization between the local system observables. Finally, we will extend our discussion of mutual synchronization to the case of three two-level atoms. 

\subsection{The case of two atoms}\label{twoatoms}

Before starting to present our results, we would like to comment on an important point that will be pertinent in the following sections. Since we intend to study the phenomenon of transient synchronization, we will be monitoring the temporal similarities between the dynamics of the expectation values of local atom observables. As a matter of fact, one can choose to focus on any physical observable of a two-level system, that can be expressed as a linear combination of the operators $\{\sigma^x,\sigma^y,\sigma^z,\mathbb{I}_2\}$, in order to discern the dynamical onset of mutual quantum synchronization. For this purpose, to be concrete in our presentation, we will be considering the expectation value of the Pauli operator in the x-direction, $\langle \sigma^x_{s_i} \rangle =\text{Tr}[\rho_i \sigma^x]$, throughout this work, where $\rho_i$ is the reduced density operator of the $i^{th}$ atomic system.

\begin{figure}[t]
\includegraphics[scale=0.85]{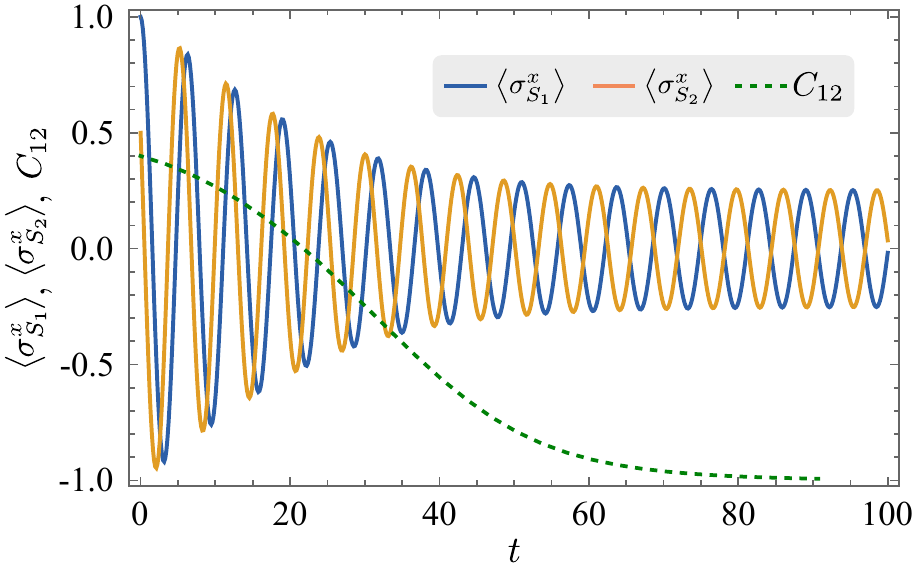}
\caption{Time evolution of the local expectation values for the atomic particles $s_1$ (dark blue line) and $s_2$ (light orange line) and the Pearson correlation coefficient $C_{12}$ (dashed green line) displaying the dynamical onset of the spontaneous mutual anti-synchronization in a two-atom system, which is collectively interacting with a zero temperature dissipative environment.}
\label{fig1}
\end{figure}

\begin{figure*}[t]
\includegraphics[scale=0.77]{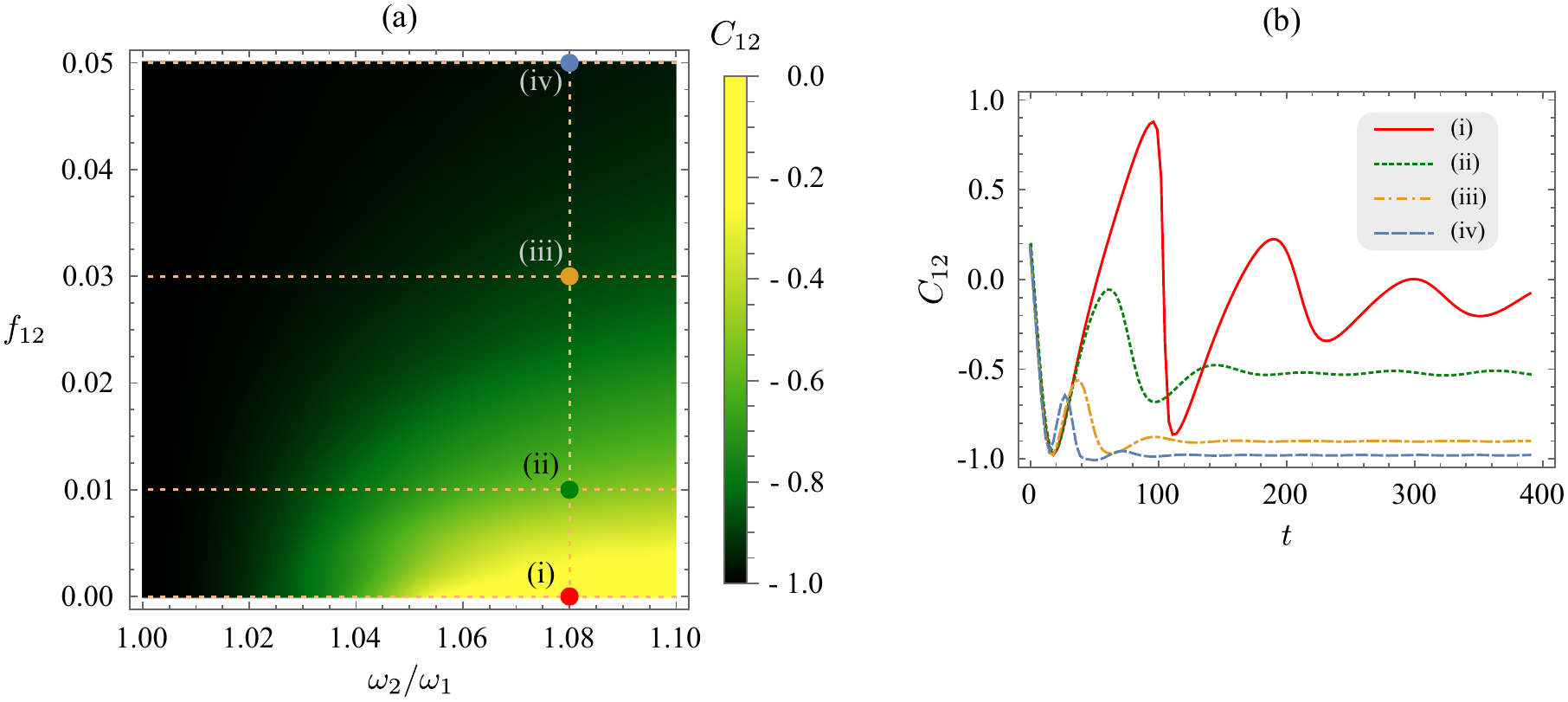}
\caption{(a) Synchronization diagram displaying the final value of the Pearson correlation coefficient $C_{12}$ with a color-coded legend, after a time interval of $t=400$, as a function of the strength of the dipole-dipole interaction between the two atoms and detuning between their self-energies. (b) Time evolution of the Pearson correlation coefficient $C_{12}$ for a set of four points, (i) red solid line, (ii) green dotted line, (iii) orange dash-dotted line and (iv) blue dashed line, which are marked in the synchronization diagram.}
\label{fig2}
\end{figure*}

Let us now begin our analysis first considering a special case of the studied model to basically comprehend the synchronizing nature of the open system dynamics in our work. Here, we assume a zero temperature environment, that is, there exists no thermally induced emission or absorption ($\bar{n}=0$). For the sake of simplicity, we also suppose that there is no detuning between the self-energies ($\omega_1=\omega_2$) and no direct interaction between the particles ($f_{12}=0$). Furthermore, we set the collectivity parameter of the dynamics as $a_{12}=1$, that is to say that we consider the case of fully collective dissipation of the two atomic particles. We should also mention that in our study, we assume the bipartite state of the two atom system to be initially uncorrelated, that is, in the form $(\cos\theta_1 |0\rangle + e^{i \phi_1} \sin\theta_1 |1\rangle)\otimes(\cos\theta_2 |0\rangle + e^{i \phi_2}\sin \theta_2 |1\rangle)$. In the present case, to be specific, we choose the initial states of the individual atomic particles as
\begin{align}
\label{init}
|\psi_{s_1}\rangle &=\cos(\pi/4)|0\rangle+\sin(\pi/4)|1\rangle \\
|\psi_{s_2}\rangle &=\cos(\pi/4)|0\rangle+e^{-i \pi /3}\sin(\pi/4)|1\rangle \nonumber
\end{align}
Nevertheless, it is important to emphasize that we have performed similar calculations for numerous random initial state pairs of the two atoms and confirmed that the outcomes of our analysis regarding synchronization remain qualitatively unchanged, despite the quantitative differences. Thus, one can say that mutual synchronization behavior does not really depend on the initial state parameters but rather imposed by the properties of the open system dynamics.

In Fig. \ref{fig1}, we display the time evolution of the expectation values for the two atomic system particles $s_1$ (dark blue line) and $s_2$ (light orange line) along with dynamics of the Pearson coefficient $C_{12}$ (dashed green line) which measures the degree of synchronization between the atoms. Here, the model parameters are fixed as $\omega_1=\omega_2=1$ and $g=0.05$, and the Pearson correlation coefficient is evaluated for a partially overlapping window $\delta t=6$ of sliding time intervals $\Delta t=9$. As can be easily observed from the plot, under the specified circumstances, the dissipation process dynamically and spontaneously gives rise to the emergence of anti-synchronized behavior of the two atoms as the Pearson coefficient $C_{12}\rightarrow -1$. Since there is no direct interaction between the system particles, the only mechanism that can lead the system into anti-synchronization is the collective interaction with the environment, which establishes an indirect interaction between the atomic systems. Therefore, the phenomenon at hand here is an example of environment-induced anti-synchronization that have also been recently observed in different models \cite{Galve2017,Karpat2019}. At this point, it is also important to note that, this phenomenon has its roots in the time-scale separation between the decay rates of the oscillations in the local observables, which are dictated by the eigenvalues of the Lindbladian governing the dynamics~\cite{Giorgi2019}. When one of these rates decays considerably slower as compared to the others such that it survives while the others quickly approach to zero, we observe the emergence of synchronous behavior between the local observables.

Having recognized the anti-synchronizing tendency of the open system dynamics in question, we are turning our attention to the possibility of detuning between the self-energies of the atomic particles $s_1$ and $s_2$. In addition to the presence of detuning, we now also allow the two particles to directly interact with each other with an exchange interaction whose strength is determined by the model parameter $f_{12}$. Once again, without any loss of generality, we suppose that the initial state of the atomic particles is described by the state pair in Eq. (\ref{init}), and $g=0.05$ and $\omega_1=1$. In Fig. \ref{fig2}(a), we present a synchronization diagram demonstrating the synchronization behavior for the two particles in terms of the detuning between their self-energies and the exchange interaction between them. Specifically, the diagram shows the final value of Pearson correlation coefficient with a color-coded legend at time $t=400$ for a partially overlapping time window $\delta t=6$ of sliding time intervals $\Delta t=9$. The reason we set this value as the final time point of our analysis is that after this time period expectation values of the observables become arbitrarily small, i.e., they vanish for all practical purposes. To put it differently, if synchronization between the atomic particles could not be established until $t=400$, this means that the open system have already reached its steady state before the emergence of transient synchronization. Looking at the diagram, one can notice that there exists a trade-off between the detuning between the atomic particles and the strength of the dipole-dipole coupling. In particular, anti-synchronized dynamics of the atomic observables ($C_{12}\approx-1$) can only be spotted when the direct exchange interaction between the two atoms is sufficiently strong to compensate for the detuning. In order to acquire a better understanding of the situation, we show the time evolution of Pearson coefficient in Fig.~\ref{fig2}(b) for four particular color-coded points marked on the diagram in Fig.~\ref{fig2}(a) with small roman numbers from (i) to (iv).
\begin{figure}[t]
\includegraphics[scale=0.75]{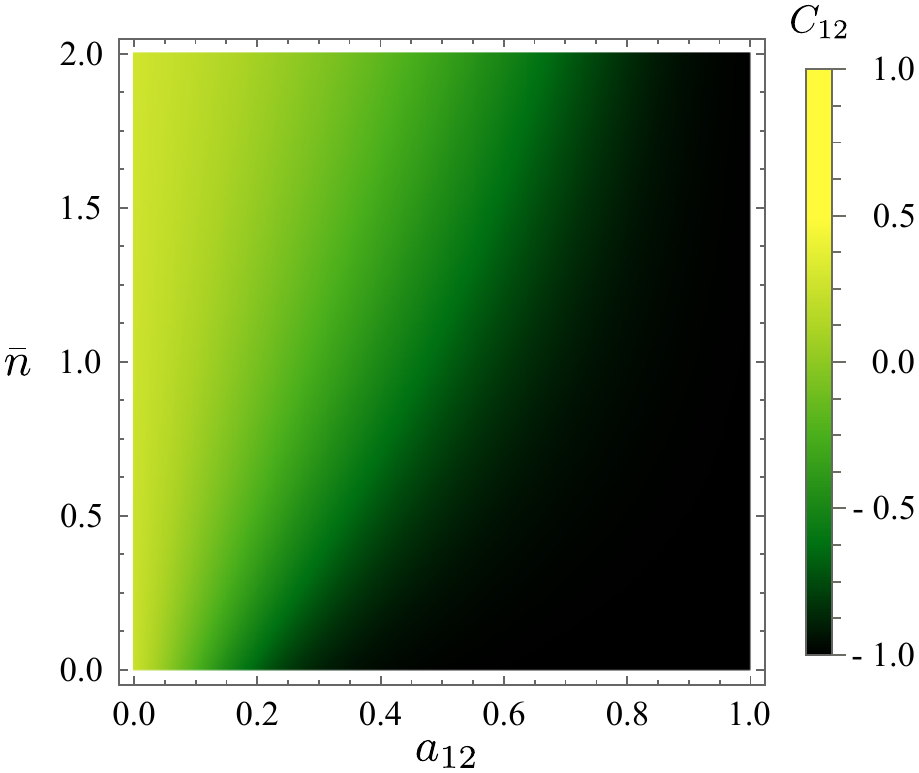}
\caption{Synchronization diagram showing the final value of the Pearson correlation coefficient $C_{12}$ with a color-coded legend, after a time interval of $t=150$, as a function of the mean number of photons and the collectivity parameter of the dissipation process.}
\label{fig3}
\end{figure}
For all of these points, detuning between the self-energies of the atoms is fixed to be $\omega_2/\omega_1=1.08$. Here we see from the plot (iv) that a direct interaction strength of $f_{12}=0.05$ completely recovers the loss of anti-synchronized evolution between the atoms, which occurs in case of vanishing exchange interaction as displayed by plot (i). When it comes to two remaining curves, namely (ii) and (iii), we observe that even though the final value of the Pearson coefficient cannot reach $C_{12}\approx-1$, it settles to a constant value within the time interval $t=400$. This in fact means that although the dynamics of the observables of the two atomic particles do not become anti-synchronized, their oscillations get dynamically phase-locked and continue to evolve with a common pace. Such a behavior is recognized in the recent literature as time-delayed synchronization and studied in relation with the collective phenomena of sub- and super-radiance \citep{Bellomo2017}. Indeed, it is also possible to say that full anti-synchronization is time-delayed synchronization simply with a phase difference of $\pi$ radians. To summarize our findings here, Fig.~\ref{fig1} and Fig.~\ref{fig2} clearly exhibit that the main role played by the direct exchange interaction between the atoms for the appearance of anti-synchronization is the compensation of the detuning between the atomic self-energies. Moreover, when the strength of this direct exchange interaction is not significant enough to induce full anti-synchronization between the two atomic particles, it might still give rise to the phenomenon of time-delayed synchronization. 

Up until this point, we have assumed that the atoms interact with a dissipative environment at zero temperature in a fully collective manner, that is, $\bar{n}=0$ and $a_{12}=1$. Therefore, the next thing we intend to do is to understand the effects of the thermally induced emission and absorption processes, and the collectivity parameter on the emergence of transient anti-synchronization. Since we have seen in our previous discussions that detuning between the self-energies of the atoms can be simply compensated with a sufficiently strong dipole-dipole interaction, here we suppose that $\omega_1=\omega_2=1$ and $f_{12}=0$ to solely focus on the impact of environment temperature and the degree of collectiveness of the dissipation on the synchronization. Thus, with the color-coded synchronization diagram displayed in Fig.~\ref{fig3}, we show the effect of the mean number of photons $\bar{n}$ at the transition frequency of the atoms and the collectivity parameter $a_{12}$, which is controlled by the distance between the atoms, on the onset of anti-synchronization. In this figure, the final value of the Pearson coefficient $C_{12}$ is evaluated at the time $t=150$ for a partially overlapping window $\delta t=6$ of sliding time intervals $\Delta t=9$. Initial state of the atomic particles $s_1$ and $s_2$ is set as in Eq.~(\ref{init}) and $g=0.05$. Our findings reveal that increasing environment temperature is detrimental for the establishment of anti-synchronized dynamics between the local observables of the atom pair. Moreover, it becomes evident that anti-synchronization cannot be achieved unless the two atoms are sufficiently close to each other. In other words, as the thermal effects in the environment become more and more significant, anti-synchronization could no longer be maintained unless a certain degree of collective dissipation is guaranteed.

\begin{figure}[t]
\includegraphics[scale=0.74]{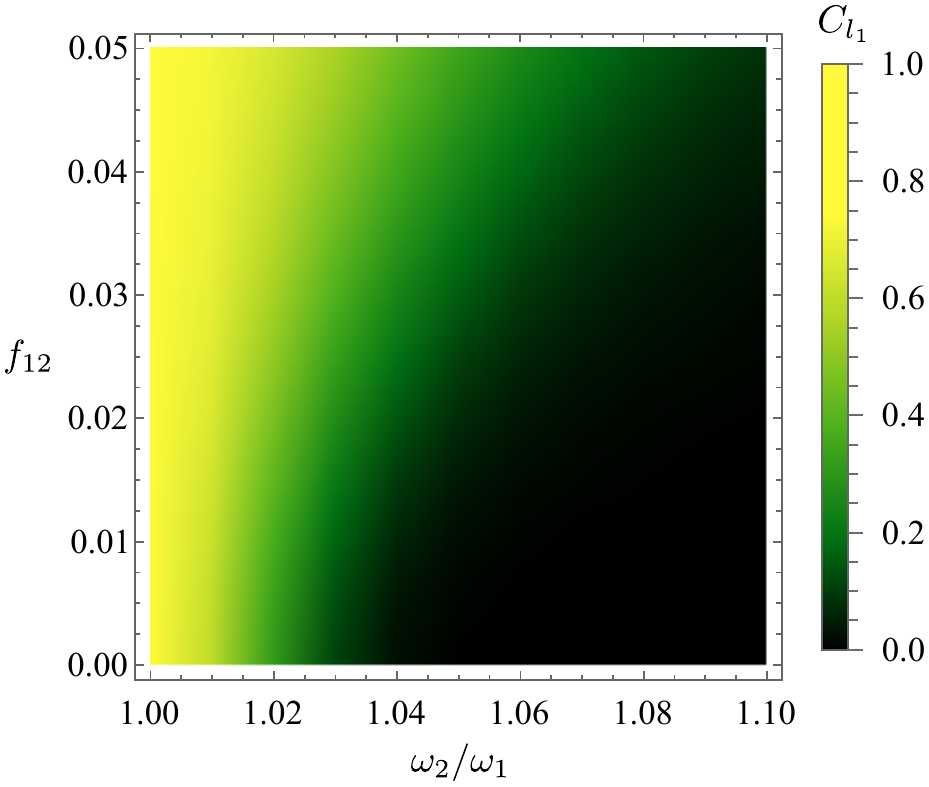}
\caption{Residual quantum coherence in the bipartite state of two atoms, as measured by $l_1$-norm coherence at time $t=400$, as a function of the strength of the dipole-dipole interaction between the two atoms and the degree of detuning between their self-energies.}
\label{fig4}
\end{figure}

In the final part of this section, we explore the existence of a possible relation between the emergence of spontaneous anti-synchronization between the expectation values of the individual atoms and the presence of quantum coherence in the two-atom system. In fact, such a link between several different quantum and total correlation measures and synchronization has been extensively investigated in the recent literature~\cite{Galve2017}. It has been understood that synchronization between the local observables can be spotted through the behavior of global correlation quantifiers only in certain models, and thus there exists no general connection between the concepts. To our knowledge, in all of the previous works on this subject, the emergence of mutual synchronization and the degree of correlations are comparatively studied after fixing the initial state of the open system. However, it is clear that the amount of correlations or coherence in the global system can be highly state dependent in general, despite the fact that the behavior of synchronization is not qualitatively affected by the choice of different initial states. In other words, quantum synchronization is evidently a property of the dynamics rather than the initial state of the system, hence, we believe that a more preferable comparison between the two concepts should be done by evaluating the average quantum coherence contained in the global system. Here, we quantify coherence in the global two-atom system using the so-called $l_1$-norm of coherence \cite{Baumgratz2014},
\begin{equation}
C_{l_1}(\rho)=\sum_{i\neq j } |\langle i | \rho |j \rangle |,
\end{equation}
which is nothing other than the sum of the absolute values of the off-diagonal elements appearing in the density operator $\rho$.

In Fig.~\ref{fig4}, we present a color-coded graph displaying the amount of average residual quantum coherence contained in the the two-atom system at time point $t=400$ as a function of the strength of the dipole-dipole interaction between the atoms and the degree of detuning between their self-energies. We have calculated the coherence values shown in the plot, averaging over $10^5$ random pairs of initial states of the form $(\cos\theta_1 |0\rangle + e^{i \phi_1} \sin\theta_1 |1\rangle)\otimes(\cos\theta_2 |0\rangle + e^{i \phi_2}\sin \theta_2 |1\rangle)$, where the real-valued angles $\theta_{1,2} \in [0,2\pi]$ and $\phi_{1,2} \in [0,\pi]$ are sampled considering a uniform probability distribution. Also, in order to allow for a direct comparison between synchronization and coherence, the model parameters here are chosen as in Fig.~\ref{fig2}, that is, $\omega_1=1$, $\bar{n}=0$ and $g=0.05$. Comparatively analyzing the diagrams in Fig.~\ref{fig2} and Fig.~\ref{fig4}, it is not difficult to observe that there is indeed a resemblance between them. In particular, one can notice that in the parameter region where the two-atom system has a relatively large residual coherence, anti-synchronized dynamics can be established between the atomic particles. In effect, for the coherence contained in the two-atom system is closely related to the expectation values of the local observables, such a link between the two might be rather expected. However, we note that this apparent relation cannot always be guaranteed in a model independent sense, since the presence of residual coherence does not always imply the manifestation of transient synchronization.

\subsection{The case of three atoms}\label{threeatoms}

In this section we ask a more curious and non-trivial question, i.e., what happens to the synchronization dynamics if we have three two-level atoms, instead of two, as our main system? Since the emergence of mutual synchronization is identified by comparatively looking at the oscillations in the local observables of two system particles, in the case of three atoms, we will need to analyze the behavior of these observables among the three possible bipartitions of the main system. We have seen in the previous section that the considered model tend to drive the atomic particles to behave in an anti-synchronous manner, and the key role in such dynamics is played by the collective interaction with the environment. Therefore, in order to keep things as simple as possible and focus on the actual cause of anti-synchronization, we will only vary the collectivity parameter in our analysis, while fixing the strength of the direct interaction between the atoms and the environment temperature to be zero, $f_{ij}=\bar{n}=0$, and we assume that all atoms are resonant, $\omega_1=\omega_2=\omega_3=1$. 

\begin{figure}[t]
\includegraphics[scale=2.5]{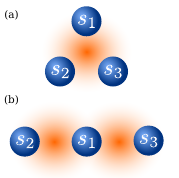}
\caption{Two different spatial configurations describing the relative inter-atomic distances for the three particle main system. (a) All of the three particles $s_1$, $s_2$ and $s_3$ are in very close proximity of each other so that they undergo fully collective dissipation. (b) The particle pairs $s_1-s_2$ and  $s_1-s_3$ are closer to each other as compared to the pair $s_2-s_3$ so that the degree of collectivity for first two pair of particles is significantly larger than the last pair. In any of these two cases, there is no direct interaction between pairs of atoms.}
\label{fig5}
\end{figure}

Fig.~\ref{fig5} schematically describes the two different spatial configurations we consider throughout this section in which we label the three atomic particles as $s_1$, $s_2$ and $s_3$. We first investigate the case of fully collective dissipation, where all system particles are very close to each other. Then, we move on to the case where all atomic particles are positioned on a line and $s_1$ is equidistant from both $s_2$ and $s_3$. Naturally, in this second case, collectivity parameters between $s_1-s_2$ and $s_1-s_3$ are identical which simply implies $a_{12}=a_{13}>a_{23}$.

The discussion of mutual synchronization for three particles already complicates things even in the simplest case depicted in Fig.~\ref{fig5}(a) where all three atomic particles experience a fully collective dissipation process. Let us briefly discuss some general aspects of this setting following a simple line of thought. We have already seen that collective dynamics in the two-atom case will lead the observables of the atoms to behave in an anti-synchronized manner. Therefore, when we bring in the third atom, one would be inclined to think that each dissipator involving any two atoms in the master equation will try to induce anti-synchronization between them. However, it is clearly not possible to achieve this; for instance, if the dynamical oscillations of the expectation value of the local observable of $s_1$ is anti-synchronized with those of $s_2$ and $s_3$, then $s_2$ and $s_3$ must in fact be fully synchronized in time. As can be already demonstrated with this simple example, the presence of a third atomic particle here could indeed dramatically affect the mutual synchronization dynamics.

\begin{figure*}[t]
\includegraphics[scale=0.85]{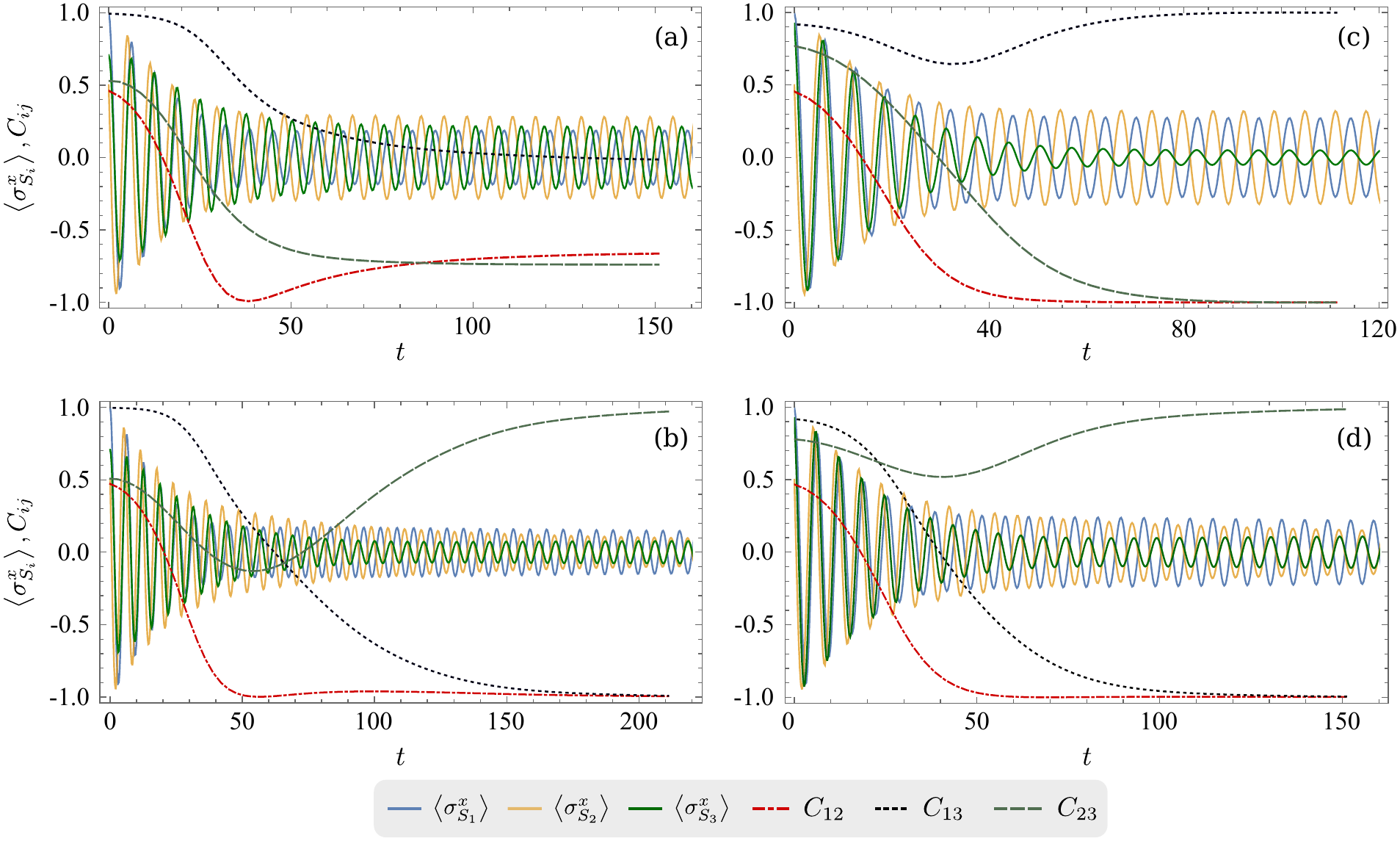}
\caption{Time evolution of the local expectation values for the resonant atoms $s_1$ (blue line), $s_2$ (light orange line) and $s_3$ (dark green line) along with the dynamics of the Pearson coefficients $C_{12}$ (red dash-dotted line), $C_{13}$ (black dotted line) and $C_{23}$ (green dashed line) between three different pairs. As the results shown in the insets (a) and (c) are for the spatial configuration given in Fig.~\ref{fig5}(a), the insets (b) and (d) display our results for the configuration described in Fig.~\ref{fig5}(b). The initial state of the first two atoms are set as $|\psi_{s_1}\rangle$ and $|\psi_{s_2}\rangle$ in all insets, and we fix the initial state of the third atom to be $|\psi_{s_3}\rangle$ in the insets (a) and (b), and to be $|\psi'_{s_3}\rangle$ in the insets (c) and (d).}
\label{fig6}
\end{figure*}

We now move on to a more quantitative discussion of the three atom case and begin the analysis of the situation at hand by studying the time evolution of the expectation values for each particle, and corresponding Pearson coefficients for different bipartitions, i.e. $C_{12}$, $C_{13}$ and $C_{23}$, presented in Fig.~\ref{fig6}. To begin with, we note that the atoms $s_1$ and $s_2$ are always chosen to be in the same initial state as in the two-atom problem throughout this subsection. We also suppose that the initial state of the third atom $s_3$ is given by
\begin{equation}
|\psi_{s_3}\rangle= \cos(\pi/8)|0\rangle+\sin(\pi/8)|1\rangle,
\end{equation}
and the Pearson correlation parameters $\Delta t$ and $\delta t$ are the same as the two-atom case.
We first assume that all three atoms are in close proximity ($a_{12}=a_{13}=a_{23}=1$), i.e., they are indistinguishable as described in Fig.~\ref{fig5}(a). The outcomes of our analysis in this case is described in Fig.~\ref{fig6}(a), where one can immediately notice that the anti-synchronous behavior observed in case of two closely situated atoms (see Fig.~\ref{fig1}) is lost and none of the atom pairs behave in a fully (anti-)synchronized manner. Additionally, it can be seen that all three Pearson coefficients $C_{ij}$ eventually settle to a particular value after a certain evolution time, which indicates a time-delayed synchronization across all bipartitions of the three particle system. At this point, we would like to emphasize that what we observe here for three atoms, undergoing fully collective dissipation, is in sharp contrast with our findings regarding the two-atom case. Thus, a few very important comments are in order. First, merely bringing in a third particle $s_3$ close to the particles $s_1$ and $s_2$, which are kept in the same initial state as in the two-atom case, might result in the loss of full mutual (anti-)synchronization. Second, in case of having two particle main system, the phenomenon of time-delayed synchronization manifests only when there is a certain degree of detuning between the self-energies of the atoms, and there also exists a non-zero direct exchange interaction between them partially compensating their detuning. However, we can see that this is no longer true when our main system is composed of three atoms since, despite the fact that all the three atoms are resonant, expectation values of the local observables of different atom pairs can still reach a common pace in time in a delayed manner. Lastly, it can be confirmed that the kind of time-delayed synchronization here cannot be compensated through enabling a sufficiently strong direct exchange interaction between the atom pairs, which is in fact what has been observed in the case of having a main system with two atomic particles as displayed in Fig.~\ref{fig2}(a).

Next, we would like to understand how the spatial configuration of the atomic particles, i.e., relative distances among them controlling the collectivity of the dissipation, affects the dynamics of mutual synchronization between pairs of atoms. For this reason, we consider the case described in Fig.~\ref{fig5}(b). In particular, we suppose that the atoms are aligned on a line in such a way that the collectivity parameters between $s_1-s_2$ and $s_1-s_3$ are the same as $a_{12}=a_{13}=0.8$ and $a_{23}=0.4$. In Fig.~\ref{fig6}(b), we demonstrate the time evolution of the expectation values of the local observables for three atoms along with the three Pearson coefficients $C_{ij}$. It is evident that in this spatial configuration fully (anti-)synchronized behavior is recovered between all atomic pairs. Due to the relatively small separation between them, $s_1-s_2$ and $s_1-s_3$ pairs become anti-synchronized in time as witnessed by the dynamics of the Pearson coefficients, that is, $C_{12} \rightarrow -1$ and $C_{13}\rightarrow -1$. In turn, such anti-synchronized behavior of $s_1-s_2$ and $s_1-s_3$ pairs inevitably induces the emergence of full synchronization between the comparatively distant particles $s_2$ and $s_3$ as detected by the Pearson coefficient $C_{23} \rightarrow 1$. It is quite interesting that a sole adjustment of inter-atomic distances through the collectivity parameters $a_{ij}$ might indeed result in a radical change in the nature of mutual synchronization for the three-atom system. Especially, it is notable that although the expectation values of the observables of the atoms tend to always get anti-synchronized in the resonant two-atom case, here it becomes possible with the inclusion of the third atom to also observe the dynamical onset of full mutual synchronization.

Having realized the fundamental role of the spatial configuration of the atomic particles for the establishment of mutual (anti-)synchronization in the three-atom case, we now intend to explore whether the nature of synchronization depends on the choice of atomic initial states. At this point, recall that the initial state of the atom pair turned out to be irrelevant in the discussion of two-atom synchronization since all non-trivially evolving initial state pairs give the same qualitative result in relation to synchronization. Therefore, one may be inclined to expect the same outcome for the initial state dependence of three atom case as well. Let us keep the initial states of the first two atoms $s_1$ and $s_2$ the same but consider, for example, the following alternative initial state for the third atom $s_3$
\begin{equation}
|\psi'_{s_3}\rangle= \cos(\pi/4)|0\rangle+e^{-i \pi /8} \sin(\pi/4)|1\rangle.
\end{equation}
For the above given initial state, we once again study the first configuration shown in Fig.~\ref{fig5}(a), where all three atoms are in very close to each other as previously discussed for the initial state $|\psi_{s_3}\rangle$ in Fig.~\ref{fig6}(a). We present the outcomes of this study in Fig.~\ref{fig6}(c), where we can immediately notice that while full mutual synchronization emerges for the atom pair $s_1-s_3$, the remaining two atom pairs $s_1-s_2$ and $s_2-s_3$ become dynamically anti-synchronized in time. Thus, a straightforward comparison of our findings in Fig.~\ref{fig6}(a) to those in Fig.~\ref{fig6}(c) reveals the crucial dependence of the mutual synchronization phenomenon on the choice of initial states of the three atoms, in sharp contrast to the two-atom problem. In fact, we see that changing only the state of the third atom $s_3$ is sufficient to alter the dynamics of mutual synchronization across all pairs of atoms in the full collective dissipation case. On the other hand, in the results presented in Fig.~\ref{fig6}(d), we consider the dynamics depicted in Fig.~\ref{fig5}(b) with $|\psi'_{s_3}\rangle$ and observe no qualitative difference as compared to Fig.~\ref{fig6}(b) in the synchronization dynamics of the system; the observables of the atoms that are closer to each other become anti-synchronized whereas the distant pair have no option but to fully synchronize.

In light of the quantitative results we have presented above, we now would like to comment and discuss some of the interesting aspects of the three atom case. From the previous section we knew that any bipartite two-level system under collective dissipation described by the considered dynamics tends towards anti-synchronization in their local observables. As also briefly discussed in the beginning of this section, it is simply impossible to have all bipartitions in the three atom case to be anti-synchronized under the same fully collective dynamics. Therefore, as we demonstrate in Figs.~\ref{fig6}(a) and (c), all Pearson coefficients settle to a certain value, however the value that they tend to is interestingly very sensitive to the initial states of the three atomic particles. This is due to some kind of frustration we bring into the system by introducing the third particle in the way depicted in Fig.~\ref{fig5}(a); the impossibility of three atom anti-synchronization leads the system to a certain time evolution, which is dependent on the initial conditions in a subtle way. Our results for the configuration shown in Fig.~\ref{fig5}(b) further supports this conclusion. Besides the two initial states we have studied to obtain the results displayed in Figs.~\ref{fig6}(b) and (d), we actually sampled numerous different random initial states for the three atom system and observed no qualitative difference from the presented Pearson factors. This suggests that the (anti)-synchronization phenomenon in the three-atom case is underpinned by the two-atom dynamics of the studied open system model. If any pair of atoms has a higher collectivity parameter than the others, that bipartition is anti-synchronized and the remaining bipartitions are forced to arrange themselves accordingly. Nevertheless, as mentioned above, if all of the collectivity parameters are identical, then the system becomes frustrated in some sense and the synchronization behavior exhibits strong initial state dependence.

As a final remark, we should mention that we have also performed a comparative analysis of the emergence of spontaneous mutual (anti-)synchronization and the residual quantum coherence contained in the three-atom state (along with the coherence in its three possible bipartitions) after a certain time evolution. Unlike the observed relation between the two concepts in case of the two-atom system, it does not seem possible to witness such a simple link when we have more than two atoms in the system because of the strong initial condition dependence of synchronization behavior due to the above mentioned geometric effects such as frustration.

\section{Conclusions}\label{conclusion}

We have analyzed the spontaneous mutual synchronization phenomenon in the case of two and three two-level atoms that are subject to dissipation by a thermal photon bath whose collectivity can be controlled by a number of parameters dependent on the spatial distances of atom pairs. Starting with the two atom case, we have identified that collective nature of the dissipation process leads to an anti-synchronized behavior in the dynamics of the expectation values of the local observables of the atoms when both atoms have identical self-energies. While any possible detuning between the energy levels of the considered atom pair can spoil the onset of anti-synchronization, we have demonstrated that this can be compensated by a sufficiently strong direct exchange interaction between them and in this way anti-synchronization can be restored. In addition, we have observed that increasing temperature in the thermal bath also have a detrimental effect on the dynamical appearance of anti-synchronization unless the dynamics of the two-atom system is fully collective, i.e., the atoms are extremely close to each other.

Extending our main system of interest to three two-level atoms have dramatically changed the synchronization behavior. Since it is not possible to have all three bipartitions of the tripartite atomic system to behave in an anti-synchronized manner, we have observed distinct qualitative outcomes for mutual synchronization dynamics depending on the spatial configuration of the atomic particles. In particular, when we assume no detuning between the energy levels of the atoms and a zero temperature environment, and supposing that all three atoms are placed very close to each other so that their dynamics are fully collective, we see that Pearson coefficients for each atom pair settle to a single value that is highly dependent on the initial state of the system particles. We argue that such a sensitivity on the initial conditions has its roots in the frustration the system experiences. Even though the dynamics tries to push each pair of atoms to anti-synchronization, due to its impossibility, they end up having different synchronization behaviors depending on the initial conditions. Then, we analyze the case in which the atoms are aligned on a line so that the one in the middle is equally close to the other two atoms which are relatively far away from each other. In this case, we have shown that irrespective of the initial state of the three atoms, the particle in the middle shows anti-synchronous behavior with the ones placed at the edges, and naturally, the atoms at the edges become fully synchronized in this way. This proves that the synchronization behavior is in fact dominated by the underlying two-atom dynamics and when the frustration is removed, the initial state dependence disappears.

On a side note, we would like to very briefly elaborate on a possible connection between synchronization and emergent non-stationary states in dissipative dynamics~\cite{Iemini2018,Buca2019,Munos2019}. Such non-stationary states exhibit persistent coherent oscillations in the long-time limit, in contrast to a stationary steady-state, and they are signalled by a purely imaginary eigenvalue subspace in the Lindbladian spectrum. Even though it is possible to observe the presence of such states in certain limits of our model (for example see the dynamics in Fig.~\ref{fig1}), this, in general, does not imply a direct relationship between the two phenomena. The reason is due to the fact that, in this work, and most of the quantum synchronization literature, what one is interested in is the transient synchronization, that is, establishment of synchronous behaviour between the local observables at a transient time during a decay into a steady-state regardless of whether it is stationary or non-stationary. Nevertheless, in the case of stationary synchronization~\cite{Giorgi2019,Cabot2019-2}, where the subsystems stay synchronized in the long-time limit, it may be interesting to investigate the relationship between the non-stationary states and synchronization.

\vspace{-0.13cm}

\begin{acknowledgments}
G.K. is supported by the BAGEP Award of the Science Academy and the TUBA-GEBIP Award of the Turkish Academy of Sciences. \.{I}.Y.\ is supported by M\v{S}MT under Grant No. RVO 14000. B.\c{C}. is supported by the BAGEP Award of the Science Academy and by The Research Fund of Bah\c{c}e\c{s}ehir University (BAUBAP) under project no: BAP.2019.02.03.. G.K. and B.\c{C}. are also supported by the Technological Research Council of Turkey (TUBITAK) under Grant No. 117F317.
\end{acknowledgments}

\bibliography{bibliography}

\end{document}